\documentclass[runningheads]{llncs}
\usepackage{graphicx}
\usepackage{float}
\usepackage{listings}
\usepackage[hidelinks]{hyperref}
\usepackage{fontawesome5}
\usepackage{amsmath}

\begin{document}
%
\title{Unified Singular Protocol Flow for OAuth (USPFO) Ecosystem}
%
%
\author{Jaimandeep Singh\inst{1}\orcidID{0000-1111-2222-3333} \and
Naveen Kumar Chaudhary\inst{1}}
\authorrunning{J. Singh  et al.}
%
\institute{National Forensic Sciences University, Gandhinagar, India}
\maketitle              
\begin{abstract}
OAuth 2.0 is a popular authorization framework that allows third-party clients such as websites and mobile apps to request limited access to a user's account on another application. The specification classifies clients into different types based on their ability to keep client credentials confidential. It also describes different grant types for obtaining access to the protected resources, with the authorization code and implicit grants being the most commonly used. Each client type and associated grant type have their unique security and usability considerations. In this paper, we propose a new approach for OAuth ecosystem that combines different client and grant types into a unified singular protocol flow for OAuth (USPFO), which can be used by both confidential and public clients. This approach aims to reduce the vulnerabilities associated with implementing and configuring different client types and grant types. Additionally, it provides built-in protections against known OAuth 2.0 vulnerabilities such as client impersonation, token (or code) thefts and replay attacks through integrity, authenticity, and audience binding. The proposed USPFO is largely compatible with existing Internet Engineering Task Force (IETF) Proposed Standard Request for Comments (RFCs), OAuth 2.0 extensions and active internet drafts.

\end{abstract}
\section{Introduction}
OAuth 2.0 is an open standard authorization framework that enables third-party applications to obtain limited access to a web service without sharing the resource owner's credentials. It is widely adopted and supported by many companies, libraries, and frameworks, and is used to protect web APIs and grant limited access to third-party applications \cite{walls2022spring}. 

Some of the most common uses of OAuth include social media platforms, such as Facebook and Twitter, to allow resource owner's (user's) to grant third-party applications access to their personal data and resources without sharing their login credentials. Identity and Access Management (IAM) systems uses OAuth to allow limited access to resources for authenticated users. Similarly, healthcare applications use OAuth to securely share patient data between different systems and providers. This allows for better coordination of care and improved patient outcomes. Also, enterprises provide secure access to company resources and data to employees and third-party applications through OAuth \cite{richer2017oauth}.

The process by which resource owners delegate access to their resources to the third-party applications instead of sharing their login credentials is also known as "delegated authorization". The temporary tokens, also known as access tokens, are issued by the authorization server, and are used by the third-party application to access the protected resources of the resource owner (user). The access tokens are issued with a specific scope, which defines the level of access that the third-party application is granted \cite{Siriwardena2020}.

\subsection{Motivation}
The OAuth 2.0 specification's inherent flexibility and lack of built-in security features can give rise to security vulnerabilities and suboptimal practices \cite{portswigger}. Developers are burdened with the responsibility of selecting the appropriate configuration options and implementing additional security measures, which can be challenging, especially for those unfamiliar with OAuth 2.0 \cite{portswigger}, \cite{8952200}. Moreover, the specification's differentiation between public and confidential clients introduces a host of challenges, including increased complexity and variability in implementation. Each client type requires distinct authentication flows and grant types, leading to separate code bases, adding to maintenance overhead and potential inconsistencies in the system. The varying levels of trust and security requirements for public and confidential clients demand careful management to avoid security risks. In particular, the implicit grant type, which is optimized for public clients are considered inherently less secure and does not support issuance of refresh tokens \cite{rfc6749}, \cite{Parecki_2019}.

Confidential and public clients may have different lifetimes for issued tokens. As a result, token management and revocation can become challenging, requiring careful handling of expiration and renewal mechanisms for each client type to maintain system security. The specification typically advises using short lifetimes and limiting the scope of access tokens issued via the Implicit flow \cite{rfc6749}, \cite{Parecki_2019}.  Scalability concerns also arise, requiring adequate resource allocation to accommodate the differing needs of confidential and public clients, especially in rapidly growing application ecosystems. Interoperability challenges may emerge when multiple entities interpret the OAuth 2.0 standard differently or support specific client types, potentially hindering seamless integration between applications and services.

\subsection{Our Research}
In our research, we analyzed various vulnerabilities impacting the OAuth 2.0. We also studied the latest Proposed Standard Request for Comments (RFCs), OAuth extensions and active Internet Drafts proposed, updated and maintained by oauth and other working groups of Internet Engineering Task Force (IETF) for improving the security and architecture of OAuth. 

Thereafter, we proposed a \textit{Unified Singular Protocol Flow for OAuth (USPFO)} that is common for both confidential and public client applications. It would also ensure consistency and security across all implementations and help to minimize the potential vulnerabilities by reducing the variability and need for maintenance of different codebases in the OAuth implementation. 

A \textit{USPFO} is a standardized approach to implementing the OAuth protocol that takes into account various OAuth extensions and best practices to provide out-of-the-box security measures against common vulnerabilities such as client impersonation, token (or code) thefts, token manipulation and replay attacks. It also supports and incorporates use of OAuth extensions such as JWT-Secured Authorization Request (JAR) RFC 9101 \cite{rfc9101}, OAuth 2.0 Pushed Authorization Requests (PAR) RFC 9126 \cite{rfc9126}, Proof Key for Code Exchange by OAuth Public Clients (PKCE) RFC 7636 \cite{rfc7636}, Active Internet Drafts such as Demonstrating Proof-of-Possession at the Application Layer (DPoP) \cite{ietf-oauth-dpop-13}, and other best practices to improve the security of the OAuth flow. The tokens are cryptographically signed and bound to a specific audience, which helps to prevent token theft and replay attacks. The cryptographic signatures also ensure authenticity and integrity for the exchanged artifacts in the protocol flow. The \textit{USPFO} is not a guarantee to solve all the vulnerabilities, but it is a step forward to minimize them.

The paper is organized as follows: In Section \ref{sec2:litsurvey}, the background of OAuth and it's commonly known vulnerabilities are discussed. The key features of USPFO and its comparison with traditional OAuth 2.0 approach are outlined in Section \ref{sec2:features}, followed by a detailed description of the protocol flow in Section \ref{sec2:protocolflow}. The various actors and their interactions are covered in Section \ref{sec2:actorsinteractions}. Section \ref{sec2:security} lays out the threat model for USPFO. Performance evaluation of USPFO is considered in Section \ref{sec2:performance}. The paper concludes with a summary and future work in Section \ref{sec2:conclusion} and \ref{sec2:future} respectively.

\section{Literature Survey}
\label{sec2:litsurvey}
\subsection{Background}
The OAuth protocol was developed by a small community of web developers from various websites and internet services to address the common problem of enabling delegated access to protected resources \cite{aaron-website}. In October 2007, OAuth 1.0, the initial stable release of the OAuth protocol, was made available. OAuth 1.0 revision A, which was released in June 2009, sought to strengthen the protocol's security and remove some ambiguities from the first iteration \cite{rfc5849}.

\subsection{OAuth 2.0}
OAuth 2.0, the current version of the OAuth protocol, was published as an RFC 6749 in 2012 \cite{rfc6749}. OAuth 2.0 introduced several significant changes from OAuth 1.0, including a simpler, more flexible architecture and a focus on providing access tokens rather than sharing login credentials. It built upon the OAuth 1.0 deployment experience but it is not backwards compatible with OAuth 1.0. 

OAuth 2.0 also introduced the concept of "bearer tokens",  [Bearer Token Usage as RFC 6750 \cite{rfc6750}] which allow third-party applications to access resources without having to authenticate the user.
\subsubsection{Grant Types}
OAuth supports four different grant types, each suited to a different client application type. The two most commonly used grant types are:

\textit{(A) \textbf{Authorization Code Grant}:} This grant type is used by \textit{confidential} clients such as server-side web applications. The confidential clients are capable of maintaining the confidentiality of their client credentials (such as a client secret). There are three steps in the authorisation code grant type. The user is first redirected by the authorization server to grant access to the third-party (client) application through User Agent (typically a web browser). The user is then redirected back to the client application with an authorization code. The client application can exchange the authorization code for an access token by sending it to the authorization server along with its client credentials \cite{richer2017oauth}.

\textit{(B) \textbf{Implicit Grant}: }This grant type is similar to the authorization code grant type, but it does not require the third-party application to exchange the authorization code for an access token. Instead, the access token is returned directly to the third-party application, which can then use it to access the protected resources of the resource owner. This grant type is optimized for \textit{public} clients, which do not have the ability to keep a secret, such as in a JavaScript-based application running in a browser, or in a native mobile app. Public clients are considered less secure than confidential clients that can prove their identity with a client secret \cite{carnell2021spring}.

\subsection{ Prevalent vulnerabilities}

The CVE (Common Vulnerabilities and Exposures) and CWE (Common Weakness Enumeration) are standards used to identify and classify security vulnerabilities in software and systems. CVE is a standard for identifying, disclosing and naming specific vulnerabilities in software and systems \cite{cve-defi}, while CWE provides a taxonomy of common weaknesses or software weaknesses that can lead to vulnerabilities \cite{cwe-defi}.

In this paper, we have identified and listed some of the prevalent OAuth vulnerabilities in terms of CVE and CWE, which provides a clear and consistent method of communicating the security issues \cite{hackerone}, \cite{bugcrowd}, \cite{portswigger}, \cite{OAuch-2022}, \cite{comp-analysis} and \cite{singh2022oauth}.

\textit{\textbf{Improper validation of redirect URIs}:} An attacker can exploit an insecure redirect URI by intercepting the authorization code and using it to gain unauthorized access to the user's resources. Maliciously crafted URI and various URI manipulation techniques are used by attackers to steal the authorization code. This vulnerability can be mitigated by using a secure redirect URI (i.e. using HTTPS) and validating the redirect URI to match the one registered with the authorization server CVE-2021-43777 \cite{CVE-2021-43777}, CVE-2022-36087 \cite{CVE-2022-36087}, CWE-601 \cite{CWE-601} and CWE-20 \cite{CWE-20}.

\textit{\textbf{Improper use of the state parameter}:} If the state parameter is nonexistent or a static value that never changes, the authorization server cannot determine if the request was initiated by the client that generated the request or by an attacker. This allows the attacker to launch a CSRF (Cross-Site Request Forgery) by sending them a malicious link that initiates an authorization request with the attacker's own client ID to gain unauthorized access to the user's resources CVE-2022-43693 \cite{CVE-2022-43693}, CVE-2023-24428 \cite{CVE-2023-24428}, CWE-352 \cite{CWE-352} and \cite{arshad2022practical}.

\textit{\textbf{Token leakage}:} It is a common vulnerability in OAuth implementations which can happen if an attacker intercepts the tokens (authorization code, access token or refresh token) that are being sent from the authorization server to the client or from the client to the protected resource server. This may occur if the connection between the entities is not secure, such as using an unencrypted connection (HTTP instead of HTTPS) or if the attacker is able to perform a man-in-the-middle (MITM) attack CVE-2022-32217 \cite{CVE-2022-32217}, CVE-2022-32227 \cite{CVE-2022-32227}, CVE-2022-39222 \cite{CVE-2022-39222}, CVE-2023-22492 \cite{CVE-2023-22492}, CWE-200 \cite{CWE-200}, CWE-312 \cite{CWE-312}, CWE-319 \cite{CWE-319}, CWE-532 \cite{CWE-532}, CWE-613 \cite{CWE-613}.

\textit{\textbf{Client impersonation}:} The client secrets are used to authenticate the client to the authorization server and should be kept confidential. If an attacker is successful in obtaining the client secrets, they can impersonate the client and obtain access tokens by using the client secret and authorization code. Malicious client can impersonate as a genuine client to gain additional restricted scopes from the authorization server that would otherwise not be granted to a less trustworthy application. This can give the attacker access to more sensitive user data or resources. To mitigate this vulnerability, the client secrets should be stored in a secure location and protected with encryption. Additionally, access to the client secrets should be restricted to only those who need it. It's also important to rotate the client's secret frequently CVE-2022-31186 \cite{CVE-2022-31186}, CVE-2019-5625 \cite{CVE-2019-5625}, CWE-522 \cite{CWE-522}, CWE-532 \cite{CWE-532} and CWE-922 \cite{CWE-922}.

\subsection{Latest developments}

There are several latest developments in OAuth that are being proposed and developed to improve the security and functionality of the protocol by working groups of Internet Engineering Task Force (IETF). These include new proposed standard RFCs, OAuth 2.0 extensions and active internet drafts such as JWT-Secured Authorization Request (JAR) RFC 9101 \cite{rfc9101}, OAuth 2.0 Pushed Authorization Requests (PAR) RFC 9126 \cite{rfc9126}, Active Internet Drafts such as OAuth 2.0 Demonstrating Proof-of-Possession at the Application Layer (DPoP) \cite{ietf-oauth-dpop-13} and OAuth 2.1 \cite{ietf-oauth-v2-1-07}. 

\subsection{The research impetus}

OAuth 2.0 specification differentiates between public and confidential clients and provides different grants for each type of client, which increases the complexity and variability. This also leads to issues with security and scalability as different code bases have to be maintained to cater to different client types.

Additionally, there is no existing unified architecture and approach that addresses known vulnerabilities in OAuth 2.0 \cite{singh2022oauth}. This can make it difficult for developers to implement OAuth securely and can lead to a lack of consistency across different implementations.
Furthermore, the piecemeal implementation and adoption of various standards also produces problems for the interoperability as some of the newer OAuth extensions and RFCs are not backward compatible, which can discourage its widespread adoption. This can make it difficult for different systems and applications to work together seamlessly and limit the potential for secure deployment of the OAuth.

\section{Features of USPFO}
\label{sec2:features}
The unified flow aims to improve the security and consistency of OAuth implementations by introducing new features and best practices. Some of the highlights of the unified flow include:

\textit{(A) \textbf{Strengthening the client authentication process}:} A separate client developer or third party controlled endpoint has been introduced to ensure the integrity and authenticity of the client application. It gives flexibility to the developer to incorporate suitable security measures to identify whether the requesting client application is genuine or otherwise. This increases the security of the overall system by providing an additional layer of defense making the system more robust and resistant against token replay and client impersonation attacks. Additionally, this approach allows for easy updating and management of the client application. This approach also follows the principle of separation of duties, as the task of ensuring genuineness of the client application initiating the OAuth process is delegated to a separate endpoint. The authorization server only verifies the client assertion claims through the pre-registered endpoint.

 \textit{(B) \textbf{Deprecating basic authentication}:} One of the major features of USPFO is to eliminate the use of the \texttt{client\_secret}, which is traditionally used by confidential clients in the authorization code grant to establish their identity with the authorization server using basic authentication method RFC 7617 \cite{rfc7617}. The \texttt{client\_secret} is generated by the authorization server at the time of registration of the client application by the developer and is shared with the client. The \texttt{client\_secret} is sent to the authorization server in base64 encoded form in the authorization code flow grant which makes it susceptible to interception through man-in-the-middle attacks. Once the \texttt{client\_secret} is compromised, it can be used to impersonate the client, making it a security risk. As the \texttt{client\_secret} is shared between the client and the authorization server, it becomes difficult to rotate the \texttt{client\_secret} without updating both the client application and the authorization server and could be a problem as it cannot be easily replaced to prevent its further unauthorized usage \cite{CVE-2022-31186}, \cite{CVE-2019-5625}, \cite{CWE-522}, \cite{CWE-532}, \cite{CWE-922}.

\textit{(C) \textbf{Merging of authorization code grant and the implicit grant}:} This feature eliminates the distinction between the two grants and unifies the architecture and protocol flow, making it easier for developers to implement, maintain and update OAuth securely.

\textit{(D) \textbf{Introducing \texttt{assertion\_verification\_uri} field}:} This new field needs to be registered at the time of registration of the client application with the authorization server. Traditionally, the \texttt{client\_secret} is used to authenticate the client to the authorization server, but it can be easily leaked or stolen, which can lead to client impersonation. The client assertion verification URI is a developer or third-party provided URI that can be used by the authorization server to retrieve the client’s public certificate and use it to verify the signature of the assertion. This also allows for easy management and rotation of keys for the client applications without the need to update the data registered with the authorization server.

\textit{(E) \textbf{In-built support for integrity, authenticity, and audience binding}:} These features provide protection against commonly known vulnerabilities such as token leakage and replay attacks. Integrity ensures that the tokens are not tampered with during transit, authenticity ensures that the tokens were issued by a trusted source, and audience binding ensures that the tokens are intended for a specific audience. Together, these features ensure that the tokens exchanged during the OAuth flow are legitimate and cannot be used by an attacker to gain unauthorized access to the protected resources.

\textit{(F) \textbf{Authentication of Client before user interaction}:} The Unified Flow (USPFO) enables authentication of client applications before any user interaction takes place as in OAuth 2.0 Pushed Authorization Requests (PAR) RFC 9126 \cite{rfc9126}. This increases the confidence in the client application for subsequent steps in the protocol flow and ensures that the request is coming from a legitimate source before any user interaction happens. This also helps in mitigation of certain kinds of attacks such as denial-of-service as the rogue requests are rejected at the initial stages itself, limiting the allocation and consumption of resources from the authorization server. 

\textit{(G) \textbf{Compatibility with newer proposed standard RFCs and OAuth extensions}:} The unified flow is largely compatible with many newer proposed standard RFCs and OAuth extensions such as JWT-Secured Authorization Request (JAR) RFC 9101 \cite{rfc9101}, OAuth 2.0 Pushed Authorization Requests (PAR) RFC 9126 \cite{rfc9126}, Proof Key for Code Exchange by OAuth Public Clients (PKCE) RFC 7636 \cite{rfc7636} which improves the security and functionality of the OAuth protocol.

\textit{(H) \textbf{A Unified Approach for Confidential and Public Clients}:} USPFO provides a unified approach for both confidential and public client applications in OAuth. A standout feature is the introduction of a separate endpoint for client authentication, eliminating the need for storing the sensitive \texttt{client\_secret}. Traditionally, confidential clients relied on the \texttt{client\_secret}, which required secure storage. USPFO's separate endpoint empowers public clients to utilize the same secure mechanism, leveling the playing field with confidential clients. This significant change removes the requirement of storing the \texttt{client\_secret} altogether, enhancing the security and parity between both client types. In-built security features of unified flow ensures integrity and authenticity, benefiting all clients. Additionally, it also unifies the two primary grant types - authorization code and implicit grants - into a single flow. This simplifies the protocol and allows developers to use a consistent approach for both confidential and public clients.

Comparison between traditional OAuth 2.0 flow and USPFO is given at Table \ref{tbl:uspfo-table}.

\begin{table*}[ht!]
  \caption{Comparison between OAuth 2.0 and USPFO}
  \label{tbl:uspfo-table}
  \includegraphics[width=\linewidth]{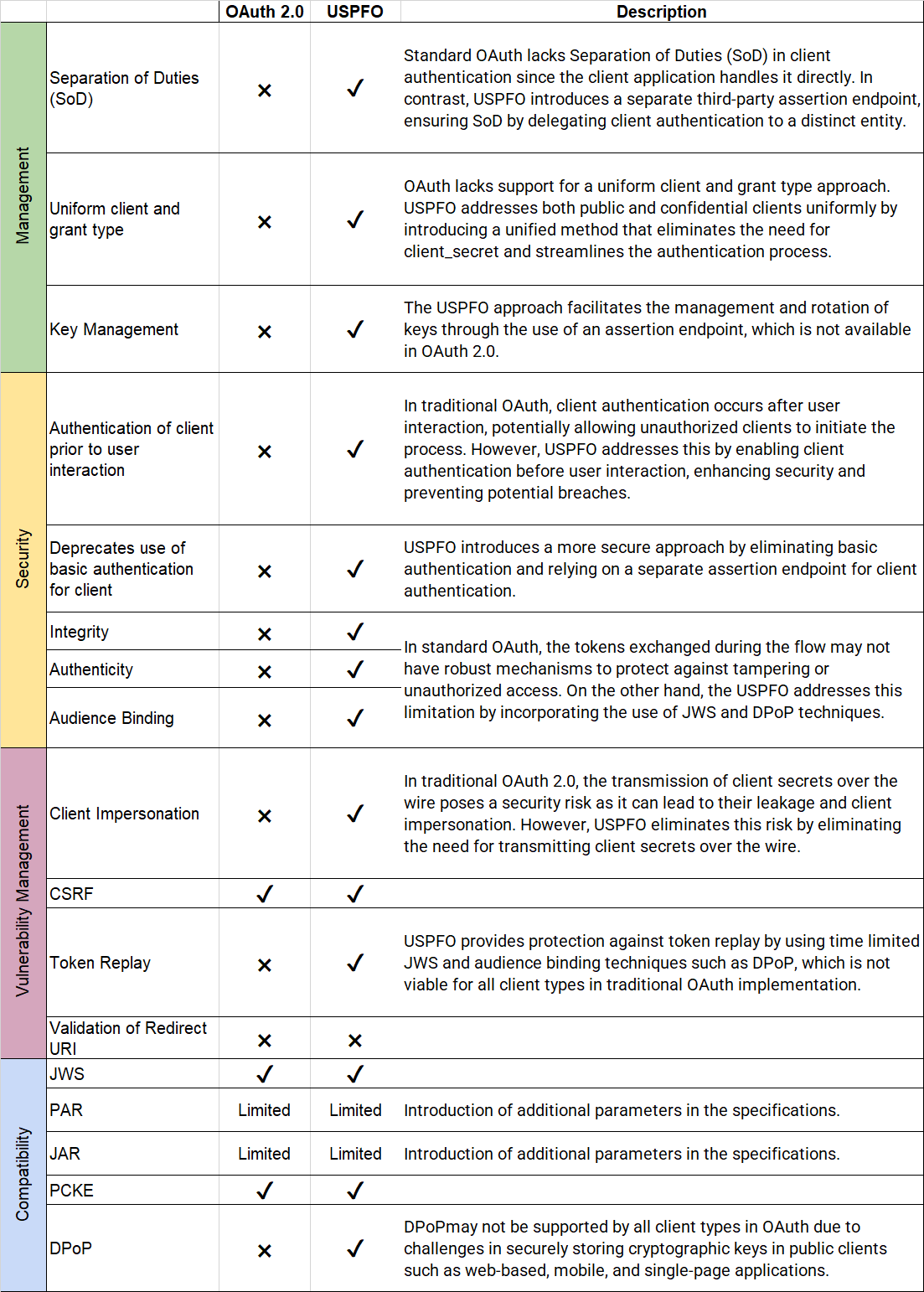}
\end{table*}

\section{Entities and The Protocol Flow}
\label{sec2:protocolflow}
\subsection{Entities Involved}
\begin{figure}[H]
	\centering
		\includegraphics[width=0.6\textwidth]{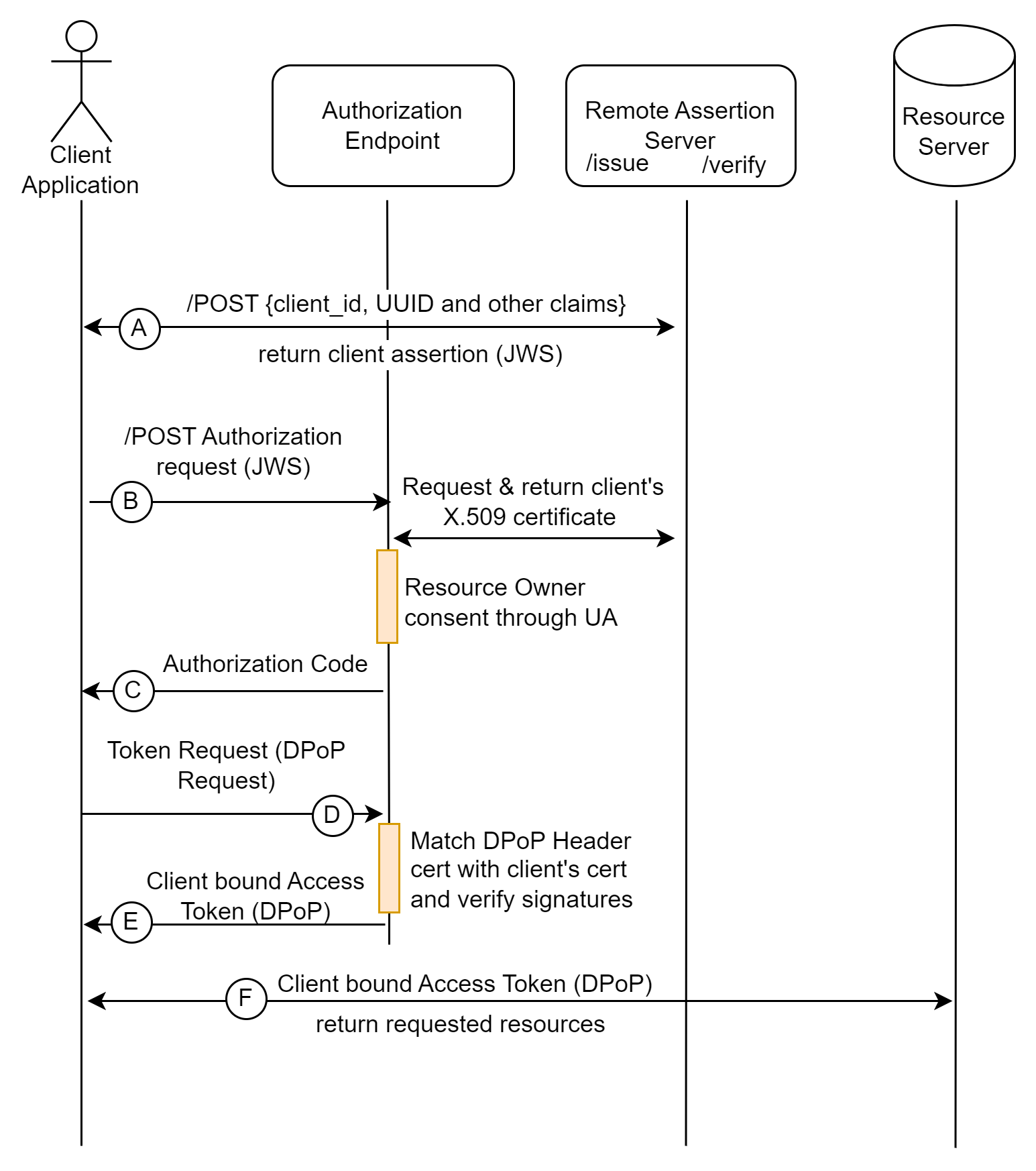}
	\caption{USPFO data and protocol flow}
	\label{FIG:uspfo}
\end{figure}

The USPFO protocol involve several key entities that play vital roles in the authentication and authorization process. In this section, we will explore the functions and interactions of the client application, remote assertion server (unique to USPFO), authorization server, and protected resource server. We will also highlight the introduction of the remote assertion server in USPFO and its significance in enhancing security and authentication for client applications. Figure \ref{FIG:uspfo} outlines the entities, their interactions, data, and protocol flow for USPFO. This section provides a comprehensive description of these components, illustrating how they collaboratively function in the USPFO ecosystem.

(i) \textbf{Client Application: }
In both OAuth 2.0 and USPFO, the client application plays a central role as it seeks access to the user's protected resources. It interacts with other components of the system to obtain the necessary permissions from the user.

(ii) \textbf{Remote Assertion Server (USPFO): }
USPFO introduces a novel addition in the form of the remote assertion server. This separate developer or third-party controlled endpoint ensures the integrity and authenticity of the client application. It enhances security, providing an additional layer of defense against client impersonation attacks. The separation of duties principle is followed as the task of verifying the client application's genuineness during the OAuth process is delegated to the remote assertion server. This design allows for easy updating, management, and incorporation of suitable security measures to identify genuine client applications.

(iii) \textbf{Authorization Server: }
The authorization server, a fundamental component in OAuth 2.0 and USPFO, plays a crucial role in verifying the client's assertions through the pre-registered endpoint provided by the remote assertion server. It ensures that the client application has the required permissions to access the user's protected resources and handles the issuance of access tokens once authentication and authorization are successfully completed.

(iv) \textbf{Protected Resource Server: }
The protected resource server is responsible for hosting and safeguarding the user's resources. It verifies the access tokens provided by the client application and allows or denies access to the requested resources based on the permissions granted during the OAuth process.

\subsection{The Protocol Flow}

The protocol flow, illustrating the interactions between various entities, is depicted in Figure \ref{FIG:uspfo} and described below:

(A) The client application sends a request having client identity (\texttt{client\_id}) and a random number [version 4 Universally Unique Identifier (UUID) string according to RFC 4122 \cite{rfc4122}] and other claims such as the intended audience to the remote assertion server. The remote assertion server will then return a signed  JWT [RFC 7519 \cite{rfc7519}] using the private key associated with the client. JSON Web Signature (JWS) is specified in RFC 7515 \cite{rfc7515} and is compact, URL-safe means of representing digitally signed or MACed content using JSON data structures. By using a remote assertion server, the client can avoid the need to store private keys on the device and it can also share the same assertion across multiple devices. 

(B) The client application sends the authorization request along with the client assertion to the authorization server.  The authorization server verifies the authenticity of the client assertion by fetching the client’s X.509 certificate from the client assertion verification endpoint, which has been registered with the authorization server at the time of registration of client application by the developer.

(C) If the client assertion is valid, the authorization server will forward the request to the resource owner through a user agent (typically a browser). If the resource owner has granted access, the authorization server will generate an authorization code and send it to the client application via the redirect URI registered with the authorization server.

(D) The client application will generate the DPoP JWT having claims as defined in the active internet draft for DPoP \cite{ietf-oauth-dpop-13}. The client application will send the DPoP JWT to the remote assertion server. The remote assertion server will sign the JWT (JWS) and return the same to the client. Additionally, it could also add the public key associated with the client as required by the DPoP specifications. 
The client will send the DPoP proof JWT in the DPoP HTTP header for the access token request. 

(E) The authorization server will fetch the client’s X.509 certificate from the remote client assertion verification endpoint and then validate the DPoP proof JWT by checking the signature, the claims, and the public key bound to the client's identity. If the DPoP proof JWT is valid, the authorization server will issue an access token (optionally refresh token) that is bound to the public key included in the DPoP proof JWT. This ensures that the access token can only be used by the client that possesses the private key associated with the public key bound to the client's identity.

(F) After the client has received the access token from the authorization server, it can use the access token to request protected resources from the resource server. The resource server checks the signatures and the validity of the access tokens and if these checks fail or the data in the DPoP proof is wrong, the server refuses to serve the request.

\subsection{Key Components of USPFO Ecosystem}
Unified Singular Protocol Flow for OAuth (USPFO), ecosystem can be summarized in following equation:

\begin{align*}
    \texttt{USPFO =} & \texttt{ assertion\_URI + JWS + pre\_authenticate\_client\_app} \\
    & + \texttt{PKCE + DPoP - basic\_auth}
\end{align*}

Here, each component plays a crucial role in enhancing the security and functionality of the OAuth protocol:

\begin{itemize}
  \item \(\textbf{assertion\_URI}\) represents the remote assertion endpoint, introduced in USPFO, which ensures the integrity and authenticity of the client application, enhancing security and protection against client impersonation attacks.
  
  \item \(\textbf{JWS}\) (JSON Web Signature) is utilized to sign the client assertions, providing integrity and authentication of the tokens exchanged during the protocol flow.
  
  \item \(\textbf{pre\_authenticate\_client\_app}\) involves the client application being authenticated prior to user interaction, increasing confidence in the application's legitimacy and security during subsequent protocol steps.
  
  \item \(\textbf{PKCE}\) (Proof Key for Code Exchange by OAuth Public Client) RFC 7636 \cite{rfc7636} is employed to protect authorization codes from interception and replay attacks, further securing the authentication process.
  
  \item \(\textbf{DPoP}\) (Demonstrating Proof-of-Possession at the Application Layer) \cite{ietf-oauth-dpop-13} is used to ensure that the client possesses the cryptographic key corresponding to the presented access token, bolstering the security of token exchange.
  
  \item \(\textbf{basic\_auth}\) USPFO deprecates the use of basic authentication for client authentication, removing a potential security vulnerability present in the traditional OAuth flow.
\end{itemize}

Together, these components form the unified approach of USPFO, enhancing security, reliability, and consistency in OAuth implementations.

\section{Actors and their Interactions}
\label{sec2:actorsinteractions}
\subsection{Client Registration}
The client application developer needs to register the client type (confidential or public) and client redirection URI with the authorization server as specified in Section 2 of RFC 6749 \cite{rfc6749}. In the unified flow architecture we introduce an additional client type \textit{unified}. 

The addition of the \textit{unified} client type allows the authorization server to distinguish between the conventional and the unified singular flow for OAuth (USPFO). The unified flow (USPFO) requires the authorization server to fetch the client’s associated X.509 certificates from a pre-registered endpoint.

The \texttt{client\_id} generated by the authorization server for the \textit{unified} client type should be prefixed by \texttt{UFO\_}, which would help in recognizing the USPFO client. This would also enable authorization server to know which flow to use when processing the request.

In addition to registering the client type and client redirection URI, the client application developer also needs to register an additional field called \texttt{assertion\_verification\_uri} for the \textit{unified} client type. This field is a reference to a trusted source that contains the client's X.509 certificates. The URI provided should be in compliance with the format specified in RFC 3986 \cite{rfc3986}, which is the standard for URI (Unified Resource Identifier) syntax. This \texttt{assertion\_verification\_uri} is used by the authorization server to fetch the client's X.509 certificates during the \textit{unified} flow of OAuth 2.0. The client assertion verification endpoint is an important part of the authentication process and helps to ensure the confidentiality and integrity of the client's credentials.

As a backup measure, the authorization server may decide to store valid client's public keys as a JSON object containing a valid JWK Set (JSON Web Key Set) as defined in RFC 7517 \cite{rfc7517}. This is to account for situations where the specified URI as referenced in \texttt{assertion\_verification\_uri} is not available or not online. However, using a URI is the preferred method as it allows the client to rotate its certificates without the need to make corresponding changes in the authorization server data. The client can use self-signed certificates or trusted third-party signed certificates for authentication. The self-signed certificate method allows for client authentication without the need to maintain a Public Key Infrastructure (PKI) RFC 5280 \cite{rfc5280}.

\subsection{Remote Assertion Server}
The use of a remote assertion server allows client applications to avoid the need to store private keys on the device, and to share the same assertion across multiple devices. The remote assertion server generates and signs the client assertion, which is then sent back to the client application, who will include it in the authorization request to the authorization server.

The remote assertion server can also be used to generate client assertions for other OAuth extensions such as JWT-Secured Authorization Request (JAR), OAuth 2.0 Pushed Authorization Requests (PAR) RFC 9126 \cite{rfc9126} and Demonstrating Proof-of-Possession at the Application Layer (DPoP) \cite{ietf-oauth-dpop-13}.

\subsection{Client Assertion Request}

The client application generates a JWT having client identity in the form of \texttt{client\_id}, a random number [Universally Unique Identifier (UUID) based 128-bit random number as defined in RFC 4122 \cite{rfc4122}] and other claims such as the intended audience.

An example of a request that a client application would send to the remote assertion server to obtain a signed client assertion is given at Listing \ref{lst:clientreq}.

\begin{lstlisting}[caption=Client assertion request, label={lst:clientreq}, float=false]
POST /assertion/issue HTTP/1.1
Host: assertion.example.com
Content-Type: application/json

{
 "alg": "RS256",
 "typ": "JWT",
 "aud": "https://server.example.org/token",
 "client_id": "UFO_s6Bk8dRkqt3", 
 "jti": "5e5ede50-dc60-40b0-bc94-cb2115ad6820"
}
\end{lstlisting}

The request body is in JSON format and contains the following claims:

\noindent "alg": The signing algorithm used to sign the JWT. "RS256" in this case.

\noindent "typ": The type of the token, "JWT" in this case.

\noindent "aud": The intended audience of the JWT. The URI of the authorization server's token endpoint in this case.

\noindent "client\_id": The client\_id of the client application.

\noindent "jti": A unique identifier for the JWT, a UUID in this case.

On receiving a request from the client application, the remote assertion server will first determine the identity and authenticity of the client application through a predetermined method. This method can vary depending on the implementation and could include techniques such as checking for a valid digital signature or comparing a hash of the application binary / source code to a known good value. One of the methods is covered in Section \ref{sec2:performance}. 

Once the remote assertion server establishes the identity of the client application it will return a signed and Base64Url encoded JWT (JWS) [RFC 7515 \cite{rfc7515}] as a response to the client application.

The authorization server can use the client's public key to verify the authenticity of the JWT, and thus authenticate the client, before any user interaction happens. This approach allows the remote assertion server to act as an independent third party that can authenticate the client without the need for the authorization server to maintain a copy of the client's secret. This also allows for more flexibility in terms of how the client is authenticated. 

An example of the signed and Base64URL encoded JWT using RS256 as defined in JSON Web Signature (JWS) RFC 7515 \cite{rfc7515} is given at Listing \ref{lst:signed}.

\begin{lstlisting}[caption=Signed and Base64URL encoded JWT using RS256, label={lst:signed}, float=false]
eyJhbGciOiJSUzI1NiIsInR5cCI6IkpXVCJ9.eyJhbGciOiJSUzI1Ni
IsInR5cCI6IkpXVCIsImF1ZCI6Imh0dHBzOi8vc2VydmVyLmV4YW1wb
GUub3JnL3Rva2VuIiwiaXNzIjoiVUZPX3M2Qms4ZFJrcXQzIiwic3Vi
IjoiVUZPX3M2Qms4ZFJrcXQzIiwiZXhwIjoxNjIzOTAyMjEzLCJqdGk
iOiIxMjM0NTY3ODkwIn0.Ho4GQlnIzcNcCg5lwYGDihMU07fQTMMzQI
RlANgjadO_OrtD0X7s2w-fm9kXcdIYJ7RpLH-41jfJ6HKOIbgAkvS9D
RvVctUR_cw3yJCC8g1tPagrw3gFDkVMBi5R0sB7q-Tl47zwFhWU2sNE
9b2g4qNWSZSZGTdVzxdzDuEhQoQNS1a2iRjZe8euBqtqfJGPlZJs61T
1peiDcJtprlb0nGs63VX0PSWUbwT8ZouL1qdOGYiZY0UurzARxOrlhB
YDw42x9fxkkZQaKspFI-ixa_Qwm2Y2TN7rRiQ9VPym_iC3H4NGEPfFm
6Xh2Cn_of9rCfMoN8OZSYz1pJ81XkxI4A 
\end{lstlisting}

\subsection{Certificate Request}
The authorization server requests the X.509 certificate associated with the \texttt{client\_id} from the remote client assertion verification endpoint. An example of the request message sent by the authorization server to the client assertion verification endpoint is at Listing \ref{lst:certrequest}. The request is sent as a HTTP POST to the endpoint's URI, such as  \texttt{/cert} in this example. The host header is set to the domain of the client assertion verification endpoint, in this case \texttt{assertion.example.org}. The Content-Type header is set to \texttt{application/x-www-form-urlencoded} and the request body contains the \texttt{client\_id} of the client, prefixed by \texttt{UFO\_} as previously described. This request is used to fetch the client's X.509 certificate from the client assertion verification endpoint, which the authorization server uses to verify the client's identity and ensure the authenticity of the request.

\begin{lstlisting}[caption=Certificate request, label={lst:certrequest}, float=false]
 POST /cert HTTP/1.1
 Host: assertion.example.org
 Content-Type: application/x-www-form-urlencoded

client_id=UFO_s6Bk8dRkqt3
\end{lstlisting}

\subsection{Authorization Request}

Once the client has obtained the signed and Base64URL encoded JWT (client assertion) from the remote assertion server, it can initiate a POST authorization request with the client assertion claim as defined in OAuth 2.0 Pushed Authorization Requests (PAR) RFC 9126 \cite{rfc9126}.
The client sends the client assertion claim in the request body, typically as a \texttt{x-www-form-urlencoded}, along with other required parameters such as the response\_type, scope, and redirect\_uri.
An example of such a request is at Listing \ref{lst:certreq}.

\begin{lstlisting}[caption=Certificate request, label={lst:certreq}, float=false]
POST /as/ufo HTTP/1.1
 Host: as.example.com
 Content-Type: application/x-www-form-urlencoded

 response_type=code
 &state=af0fijsdlkj
 &client_id=UFO_s6Bk8dRkqt3
 &redirect_uri=https%3A%2F%2Fclient.example.org%2Fcb
 &code_challenge=
 GjuFoFczD6KdsLNRpqtbv0dOlGGLUNEX6WTRSAnIZFc
 &code_challenge_method=S256
 &scope=read%20write
 &client_assertion_type=
 urn%3Aietf%3Aparams%3Aoauth%3Aclient-assertion-type%
 3Ajwt-bearer
 &client_assertion=
 eyJhbGciOiJSUzI1NiIsInR5cCI6IkpXVCJ9.eyJhbGciOiJSUzI1
 NiIsInR5cCI6IkpXVCIsImF1ZCI6Imh0dHBzOi8vc2VydmVyLmV4Y
 W1wbGUub3JnL3Rva2VuIiwiaXNzIjoiVUZPX3M2Qms4ZFJrcXQzIi
 wic3ViIjoiVUZPX3M2Qms4ZFJrcXQzIiwiZXhwIjoxNjIzOTAyMjE
 zLCJqdGkiOiIxMjM0NTY3ODkwIn0.Ho4GQlnIzcNcCg5lwYGDihMU
 07fQTMMzQIRlANgjadO_OrtD0X7s2w-fm9kXcdIYJ7RpLH-41jfJ6
 HKOIbgAkvS9DRvVctUR_cw3yJCC8g1tPagrw3gFDkVMBi5R0sB7q
 -Tl47zwFhWU2sNE9b2g4qNWSZSZGTdVzxdzDuEhQoQNS1a2iRjZe8
 euBqtqfJGPlZJs61T1peiDcJtprlb0nGs63VX0PSWUbwT8ZouL1qd
 OGYiZY0UurzARxOrlhBYDw42x9fxkkZQaKspFI-ixa_Qwm2Y2TN7r
 RiQ9VPym_iC3H4NGEPfFm6Xh2Cn_of9rCfMoN8OZSYz1pJ81XkxI4
 A
\end{lstlisting}

\subsection{Authorization Response}
Once the resource owner grants the access, the authorization server issues an authorization code to the client application through redirection URI.  An example HTTP response in \texttt{x-www-form-urlencoded} is at Listing \ref{lst:authres}:

\begin{lstlisting}[caption=Authorization Response, label={lst:authres}, float=false]
 HTTP/1.1 302 Found
Location: https://client.example.org/cb?code=
SplxlQBZeOOYbYSW6xSbIA&state=af0fijsdlkj
\end{lstlisting}

\subsection{Token Request}
The client will make a request to the remote assertion server by sending the JSON data as defined in the DPoP (Demonstrating Proof-of-Possession at the Application Layer) \cite{ietf-oauth-dpop-13} header along with the client identity. The remote assertion server will use the client's private key associated with the \texttt{client\_id} to sign the DPoP proof JWT and return the signed and Base64URL encoded JWT (JWS)[RFC 7515 \cite{rfc7515}] as a response to the client application.

An example of the POST request sent by client is at Listing
\ref{lst:tokreq}.

\begin{lstlisting}[caption=Token request, label={lst:tokreq}, float=false]
POST assertion/dpop HTTP/1.1
Host: assertion.example.com
Content-Type: application/json

{
  "typ":"dpop+jwt",
  "alg":"ES256",
  "jti":"f5f254f2-15e1-4a87-9ace-e7a30601df79",
  "htm":"POST",
  "htu":"https://server.example.org/token",
   "client_id": "UFO_s6Bk8dRkqt3"
}
\end{lstlisting}

The client must provide a valid DPoP proof JWT in a DPoP header when sending an access token request to the authorization server's token endpoint in order to request an access token that is bound to a public key of the client using DPoP. The HTTP request shown in Listing \ref{lst:atokreq} demonstrates such an access token request utilizing an authorization code grant with a DPoP proof JWT in the DPoP header.

\begin{lstlisting}[caption=Access token request, label={lst:atokreq}, float=false]
POST /token HTTP/1.1
Host: server.example.com
Content-Type: application/x-www-form-urlencoded
DPoP: 
eyJ0eXAiOiJkcG9wK2p3dCIsImFsZyI6IkVTMjU2IiwiandrIjp7Imt0
eSI6IkVDIiwidXNlIjoic2lnIiwiY3J2IjoiUC0yNTYiLCJ4IjoiZW5W
UXVRWjdmN1k4SFFKdWtqamVCaFJHaWY0VzA2TlhDdk5ENWlVams4ayIs
InkiOiJwMXllUndsN3R5YlBwZWoxbDVkNWhSdUg2V252N01RMDZQbmho
UDVrR2s4IiwiYWxnIjoiRVMyNTYifX0.eyJqdGkiOiJmNWYyNTRmMi0x
NWUxLTRhODctOWFjZS1lN2EzMDYwMWRmNzkiLCJodG0iOiJQT1NUIiwi
aHR1IjoiaHR0cHM6Ly9zZXJ2ZXIuZXhhbXBsZS5vcmcvdG9rZW4iLCJp
YXQiOjE2NzQzODY3Njd9.YZBDPYwgznwajvZrCyJB3Z0ECDeIbA_3su9
HOrSXhE9yXT2-aVpsp5RhQ2VMCArQWFqvfIEfawMz5y90yi97JQ

grant_type=authorization_code
&client_id=UFO_s6Bk8dRkqt3
&code=SplxlQBZeOOYbYSW6xSbIA
&redirect_uri=https%3A%2F%2Fclient%2Eexample%2Eorg%2Fcb
&code_verifier=X3ERFdPnKTKk7aUO3_X87SvykRINAXXwKSOlymaqAn
FTb4hjVE4KVzXgeyNbk06SqIC5G4_2zrcoqZ1cF4nz0-beYe04iPdDQ79
EmdcOo3Zajo08oQaXi2R5V8Z3Bk5D

\end{lstlisting}

In OAuth 2.0 [RFC 6749 \cite{rfc6749}], confidential clients are required to authenticate with the authorization server and are typically issued client passwords at the time of registration of the client application by the developer with the authorization server. These \texttt{client\_id} and \texttt{client\_secret} are used to authenticate client application using basic HTTP authentication. However, in USPFO, client passwords must not be used for client authentication. Instead, client authentication is carried out through the use of a public/private key pair and the \texttt{assertion\_verification\_uri}. This eliminates the need to store \texttt{client\_secret} and eliminates the risk of \texttt{client\_secret} being compromised due to insecure storage or intercepted over the wire. Additionally, by using public key cryptography, the client can prove possession of the associated private key. This makes it more secure as compared to basic authentication which only verifies the \texttt{client\_secret}.

To sender-constrain the access token, the authorization server associates the issued access token with the public key from the DPoP proof. This ensures that the access token can only be used by the client that requested it, and not by any other party. An example response is at Listing \ref{lst:atokres}.

\begin{lstlisting}[caption=Access token response, label={lst:atokres}, float=false]
HTTP/1.1 200 OK
Content-Type: application/json
Cache-Control: no-store

{
 "access_token": "Kz~8mXK2aElYnzHw-LC-1fBAo.4jLp~zsPE_NeO.gxU",
 "token_type": "DPoP",
 "expires_in": 2677,
 "refresh_token": "Q..Zmk92lexi8VnWg2zPW1x-tgaG0dIbc3s3EMw_Ni4-k"
}
\end{lstlisting}

Resource servers must be able to verify that an access token is bound to a public key using DPoP and have the necessary information to confirm this binding. This could be achieved by linking the public key with the token in a way that the protected resource server can access, such as by including the JWK hash in the token directly or through token introspection.

\section{Threat Model: USPFO}
\label{sec2:security}
The threat model described below provides an overview of potential threats and corresponding countermeasures for the USPFO Ecosystem.

(A) \textbf{Threat: Availability of the Remote Assertion Endpoint}

\begin{itemize}
    \item \textbf{Threat Actor}: Malicious clients, DDoS attackers, or network failures.
    \item \textbf{Objective}: Disrupt the availability of the remote assertion endpoint to prevent client assertion issuance and verification.
    \item \textbf{Countermeasures}:
    \begin{itemize}
        \item Implement \textbf{DDoS protection mechanisms} to mitigate DDoS attacks.
        \item Use \textbf{high availability and redundancy} for the remote assertion endpoint to ensure continuous availability.
        \item Employ \textbf{rate limiting and access controls} to prevent abuse and excessive traffic.
    \end{itemize}
\end{itemize}

(B) \textbf{Threat: Theft of Signed Client Assertions}

\begin{itemize}
    \item \textbf{Threat Actor}: Attackers conducting man-in-the-middle attacks or intercepting traffic.
    \item \textbf{Objective}: Steal signed client assertions during transit, leading to client impersonation and unauthorized access.
    \item \textbf{Countermeasures}:
    \begin{itemize}
        \item \textbf{Limit the lifespan of signed assertions} to reduce the window of opportunity for theft.
        \item Use \textbf{Transport Layer Security (TLS)} for secure communication between the remote assertion endpoint and the client to prevent interception and tampering.
        \item Consider \textbf{token encryption} to protect sensitive data within the client assertions.
    \end{itemize}
\end{itemize}

(C) \textbf{Threat: Data Integrity from Client to Assertion Endpoint}

\begin{itemize}
    \item \textbf{Threat Actor}: Attackers attempting to tamper with data during transmission.
    \item \textbf{Objective}: Modify data sent from the client to the remote assertion endpoint to manipulate client assertions or insert unauthorized information.
    \item \textbf{Countermeasures}:
    \begin{itemize}
        \item Use \textbf{secure TLS connection} to ensure data integrity during transit.
        \item \textbf{Ensures data integrity} by detecting any tampering or unauthorized modifications to the data during transmission through mechanisms, such as HMAC.
    \end{itemize}
\end{itemize}

(D) \textbf{Threat: Unauthorized Access to Assertion Endpoint}

\begin{itemize}
    \item \textbf{Threat Actor}: Unauthorized users attempting to gain access to the assertion endpoint.
    \item \textbf{Objective}: Bypass authentication and gain unauthorized control over the assertion endpoint.
    \item \textbf{Countermeasures}:
    \begin{itemize}
        \item Implement \textbf{strong authentication mechanisms} to restrict access to the assertion endpoint.
        \item Enforce \textbf{access controls} and \textbf{authentication checks} for all incoming requests to the endpoint.
        \item Regularly \textbf{monitor access logs} to detect and respond to any unauthorized access attempts.
    \end{itemize}
\end{itemize}

\section{Performance Evaluation of USPFO}
\label{sec2:performance}
To evaluate the performance of the Unified Singular Protocol Flow for OAuth (USPFO), we can analyze its protocol flow using the example of a Single Page App (SPA), a public app in traditional OAuth. USPFO introduces additional steps give at Figure \ref{FIG:perform} to enhance security and to arrive at a uniform flow:

\begin{figure}[H]
	\centering
		\includegraphics[width=\linewidth]{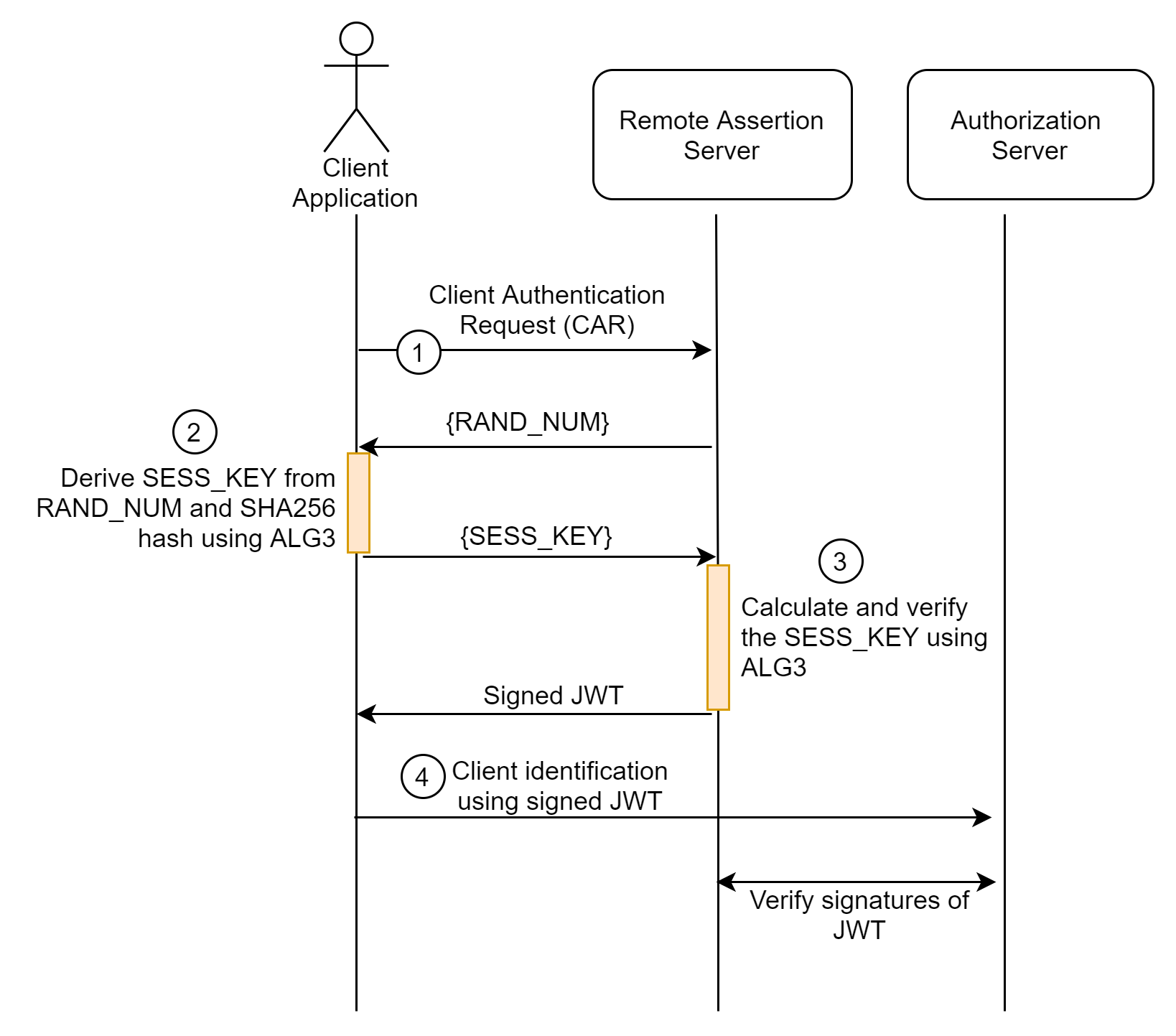}
	\caption{Performance Evaluation}
	\label{FIG:perform}
\end{figure}

\textbf{Step 1: Client Authentication Request (CAR)}

The client app sends a verification API request \texttt{(CAR)} to the Assertion Server. The server responds with a random number \texttt{(RAND\_NUM)}.

\textbf{Step 2: Session Key Generation}

The client app takes the random number \texttt{(RAND\_NUM)}, received from the assertion server, and the SHA256 hash of its source code as inputs to a proprietary algorithm(ALG3), which is specific to the application and the assertion server. The SHA256 hash ensures data integrity and security, and is commonly employed for Subresource Integrity (SRI) in Single Page Apps (SPAs) \cite{10.1145}. SRI is a security feature that allows browsers to verify that the resources loaded on a webpage, such as scripts and stylesheets, have not been tampered with or altered. It achieves this by comparing the hash of the requested resource with the expected hash provided in the webpage \cite{devdatta2016subresource}. The proprietary algorithm (ALG3) generates a session key \texttt{(SESS\_KEY)}. The \texttt{SESS\_KEY} serves as a unique identifier for the client's authentication session.

\textbf{Step 3: Session Key Verification}

The generated \texttt{SESS\_KEY} is sent back to the assertion server through an API call. The server compares the \texttt{SESS\_KEY} provided by the SPA with the key derived at its end. If they match, the server returns a signed JSON Web Token (JWT) to the client app.

\textbf{Step 4: Client Authorization}

The client app sends this JWT to the Authorization Server to identify itself. The signatures of the JWT are verified by the Authorization Server, either through Public Key Infrastructure (PKI) or the preregistered verify endpoint of the assertion server.

\subsection{Performance Overhead}
To calculate the performance overhead of USPFO, we can consider the following metrics:

(A) \textbf{API Calls ($\Theta$)}: The number of API calls made during the protocol flow, including requests and responses to various endpoints. This includes the API calls involved in client authentication and exchanging information between the client, assertion server, and authorization server.

(B) \textbf{Algorithm Complexity ($\Phi$)}: The complexity of the proprietary algorithm used for session key generation (ALG3). This complexity is determined by the computational resources required for the algorithm to generate the \texttt{SESS\_KEY} by taking inputs as the random number \texttt{(RAND\_NUM)} and the SHA256 hash of the client application's source code. A higher complexity may require more computational power and time.

(C) \textbf{Signature Verification ($\zeta$)}: The process of verifying the signatures of tokens, such as JSON Web Tokens (JWTs) exchanged during the flow. This step ensures the authenticity and integrity of the tokens. The performance impact depends on the efficiency of the signature verification process.

The performance overhead (\text{PO}) can be calculated as:

\[\text{PO} = 3 \cdot \Theta + 2 \cdot \Phi + \zeta\]

\textit{Note:} Once the JSON Web Token (JWT) expires, the client app must repeat the authentication cycle. The client app initiates a new authentication request to the assertion server, which responds with a new \texttt{RAND\_NUM}. The client app generates a new session key \texttt{SESS\_KEY} using its proprietary algorithm (ALG3) by combining the \texttt{RAND\_NUM} and the SHA256 hash of its source code. The assertion server verifies the new session key and returns a new signed JWT to the client app, allowing further interactions with protected resources.

\section{Conclusion}
\label{sec2:conclusion}
In this research paper we presented the concept of a Unified Singular Protocol Flow for OAuth (USPFO) Ecosystem. The current OAuth 2.0 framework has multiple client types and grant types, each with their own set of vulnerabilities and configuration complexities. To address these issues, we proposed USPFO which merges different client and grant types into a single, unified protocol flow. This approach not only simplifies the implementation and configuration process but also provides out-of-the-box integrity, authenticity, and audience binding for exchanged codes and tokens, protecting them against theft, replay and impersonation attacks. Additionally, USPFO is largely compatible with IETF proposed standards RFCs, OAuth 2.0 extensions and active internet drafts, such as JAR, PAR, PKCE, JWS, and DPoP. Overall, USPFO offers a solution to improve the security and usability of OAuth 2.0 for both confidential and public clients.

\subsection{Acknowledgements}
We express our gratitude to National Forensic Sciences University, Gandhinagar, India for supporting the research work and providing the necessary infrastructure and the computing resources.

\section{Future Work}
\label{sec2:future}
Future work would involve evaluating the performance and security of USPFO approach, and preparing it as a draft RFC for submission to IETF.

%
%
%
\bibliographystyle{splncs04}
\bibliography{references}
%





\begin{minipage}{0.3\textwidth}
\centering
\includegraphics[width=\linewidth]{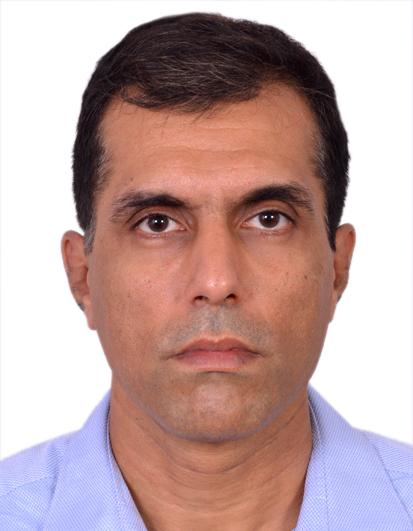}
\end{minipage}
\hfill
\begin{minipage}{0.6\textwidth}
\textbf{Jaimandeep Singh}, CISSP is a seasoned cybersecurity expert with over two decades of experience in designing, developing, and implementing IT and cybersecurity solutions. He is currently working for the Government of India in the cybersecurity field. He holds a Bachelor's and Master's degree in Computer Science and is pursuing PhD in cybersecurity. 
\href{https://www.linkedin.com/in/jaimandeep-singh-07834b1b7/}{\faLinkedin}
\end{minipage}

\begin{minipage}{0.3\textwidth}
\vspace{0.5cm}
\centering
\includegraphics[width=\linewidth]{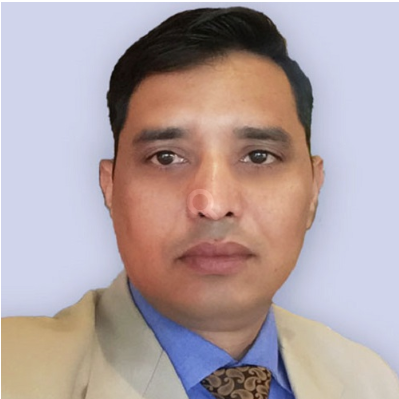}
\end{minipage}
\hfill
\begin{minipage}{0.6\textwidth}
\textbf{Naveen Kumar Chaudhary} is a Professor and Dean of School of Cyber Security \& Digital Forensics, National Forensic Sciences University, India. He has extensive experience of more than 25 years in Cyber Security, e-Governance, Digital Forensics, Network Security \& Forensics and Communication Engineering.
\end{minipage}
\end{document}